\documentclass[journal=jpcbfk,manuscript=article]{achemso}

\usepackage{graphicx}
\usepackage{amsmath}
\usepackage{amssymb}
\usepackage{amsfonts}
\usepackage{hyperref}
\hypersetup{colorlinks=true,linkcolor=blue,urlcolor=blue}
\newcommand{\be}{\begin{equation}}
\newcommand{\ee}{\end{equation}}
\newcommand{\ba}{\begin{eqnarray}}
\newcommand{\ea}{\end{eqnarray}}

\SectionNumbersOn

\author{Santi Prestipino}
\affiliation[Universit\`a di Messina]
{Dipartimento di Scienze Matematiche e Informatiche, Scienze Fisiche e 
Scienze della Terra, Universit\`a degli Studi di Messina, 
Viale F. Stagno d'Alcontres 31, 98166 Messina, Italy}
\email{sprestipino@unime.it}
\author{Domenico Gazzillo}
\affiliation[Universit\`a  di Venezia]
{Dipartimento di Scienze Molecolari e Nanosistemi, 
Universit\`a di Venezia ``Ca' Foscari'', Via Torino 155, 30172 
Venezia Mestre, Italy}
\author{Gianmarco Muna\`o}
\author{Dino Costa}
\affiliation[Universit\`a di Messina]
{Dipartimento di Scienze Matematiche e Informatiche, Scienze Fisiche e 
Scienze della Terra, Universit\`a degli Studi di Messina, 
Viale F. Stagno d'Alcontres 31, 98166 Messina, Italy}

\title{Complex Self-Assembly from Simple Interaction Rules
in Model Colloidal Mixtures}

\begin{document}
\begin{abstract}
Building structures with hierarchical order through the self-assembly of smaller blocks is not only a prerogative of nature, but also a strategy to design artificial materials with tailored functions. We explore in simulation the spontaneous assembly of colloidal particles into extended structures, using spheres and size-asymmetric dimers as solute particles, while treating the solvent implicitly. Besides rigid cores for all particles, we assume an effective short-range attraction between spheres and small monomers to promote, through elementary rules, dimer-mediated aggregation of spheres. 
Starting from a completely disordered configuration, we follow the evolution of the system at low temperature and density, as a function of the relative concentration of the two species. When spheres and large monomers are of same size, we observe the onset of elongated aggregates of spheres, either disconnected or cross-linked, and a crystalline bilayer. As spheres grow bigger, the self-assembling scenario changes, getting richer overall, with the addition of flexible membrane sheets with crystalline order and monolayer vesicles. With this wide assortment of structures, our model can serve as a viable template to achieve a better control of self-assembly in dilute suspensions of microsized particles.
\end{abstract}

\section{INTRODUCTION}

Various biomolecules, like phospholipids, peptides, and DNA filaments, as well as many synthetized colloidal particles, have the capability of assembling into mesophases, owing to their chemical and structural versatility (see, for instance, Ref.~\citenum{Bianchi}). The spontaneous assembly of colloidal particles into extended structures, like gels or membranes, is an emergent phenomenon of utmost importance in the design of functional materials. One motif that may serve different purposes is a colloidal sphere endowed with one or more attractive caps, so-called ``patches'', obtained by grafting appropriate functional groups to the sphere surface --- see examples in Refs.~\citenum{Kern,Zhang2,Russo,Munao1}. When assembled in a connected network characterized by a high surface-to-volume ratio, patchy particles may provide a practical morphology for nanoporous catalysts.~\cite{Fechete} At lower densities, Janus particles form micelles and even small bilayer shells.~\cite{Sciortino} Surfactants, i.e. molecules with an amphiphilic character, are another class of substances producing micelles in water. By forming micelles, surfactant molecules avoid the contact of their hydrophobic groups with water, thereby minimizing distortion of the hydrogen-bond network. Surfactants may also self-assemble into vesicles (closed bilayers).~\cite{Holmberg} Recently, there has been growing interest in vesicles because of their wide application in biology and medicine as model cell membranes, and for their potential as drug carriers and encapsulating agents.~\cite{Yoon,Tang,Munao2} Vesicles can also be shaped with lipids (``liposomes'') and block copolymers (``polymersomes'').~\cite{Zhang} Whether micelles or vesicles are formed depends on a subtle balance between entropy and energy. While entropy always favors spherical micelles, energetic/packing considerations put restrictions on the size and shape of aggregates: single-chain amphiphiles tend to form globular or rod-like micelles, whereas double-chain molecules prefer making bilayers.~\cite{Israelachvili} Naturally, crystallization is a simpler form of self-assembly. A large variety of complex crystals and quasicrystals has recently been obtained using particles with anisotropic shape~\cite{Damasceno} or isotropic interactions featuring multiple potential wells.~\cite{Engel} The wealth of supramolecular structures in materials with directional interactions provides the original motivation for seeking theoretical models that can be employed for a bottom-up description of these systems.

Our challenge is to obtain a complex phase behavior with minimal assumptions on the interparticle forces, possibly without modifying the interaction laws in response to a change in the target structure. In this respect, we have recently ascertained the usefulness of size-asymmetric dimers as encapsulating agent for spherical particles in a colloidal-poor solution.~\cite{Munao3,Prestipino1} Inspired by those findings, we provide in this paper a systematic study of aggregation in model colloidal mixtures of spheres and dimers. A rich self-assembly diagram emerges in the low-density regime, which counts many diverse aggregates as a function of concentration and size unbalance between the species --- including a gel-like network, a crystalline bilayer, various shapes of crystalline membranes, and spheroidal vesicles.

\section{MODEL AND METHOD}
Within an implicit-solvent scheme, we consider a dispersion of two colloidal species: a sphere (A) and a dimer (B) made up of two tangent spherical monomers --- one end (B$_1$) being three times smaller than the other (B$_2$). The A particle is represented as a hard sphere of diameter $\sigma_{\rm A}=d\sigma_{\rm B_2}$ with $d=1,2$, or 3 in this work. All particle interactions are hard-core with additive diameters $\sigma_{\alpha\beta}=(\sigma_\alpha+\sigma_\beta)/2$, except for the A-B$_1$ interaction, given by a hard-core plus square-well potential:
\ba
u(r)=
\begin{cases}
\infty & \quad \text {for} \quad r<\sigma_{\rm AB_1} \\[4pt]
-\epsilon & \quad \text{for}  \quad \sigma_{\rm AB_1}\le r\le\sigma_{\rm AB_1}+\sigma_{\rm B_1}\\[4pt]
0 & \quad \text{otherwise}\,.
\end{cases}
\label{eq:potential}
\ea
With such interaction rules, at low density and temperature spheres get coated with dimers. The rather strong asymmetry in size between B$_1$ and B$_2$ ensures a more effective encapsulation of spheres by dimers. No mutual attraction is assumed between two dimers or between two spheres, with the idea that such interactions (which are usually present in real colloids~\cite{Wolters}) are much weaker than $\epsilon$. In this way we keep the system as simple as possible in order to identify the minimal ingredients for a fairly complex self-assembly diagram. The size difference between A and B$_2$, expressed by the ratio $d$ of their diameters, is the only free parameter left in the model. We take $\sigma_{\rm B_2}$ and $\epsilon$ as units of length and energy respectively, in turn defining a reduced distance $r^*=r/\sigma_{\rm B_2}$ and a reduced temperature $T^*=k_{\rm B}T/\varepsilon$ ($k_{\rm B}$ being the Boltzmann constant); hereafter, reduced units are assumed, omitting asterisks altogether. Finally, we denote by $N_{\rm A}$ and $N_{\rm B}$ the number of spheres and dimers, respectively. Hence $N=N_{\rm A}+N_{\rm B}$ is the total number of particles and $\chi=N_A/N$ is the concentration of spheres.

We perform Monte Carlo (MC) simulations of the A-B mixture in the canonical ensemble, using the standard Metropolis algorithm with periodic conditions at the boundaries of a cubic box. Canonical conditions mimic the natural setting of a mixture of spheres and dimers where the number of particles of each species and the volume of the container are fixed. One MC cycle consists of $N$ Metropolis moves. For dimers, one trial move is a random choice between a center-of-mass translation and a rotation about a coordinate axis. The maximum shift and rotation are adjusted during the first part of the run so that the ratio of accepted to total number of moves stays close to 60\%. The acceptance rule as well as the schedule of the moves are so designed to satisfy detailed balance.

%
%
\begin{table}[t]
\begin{center}
\caption{Thermodynamic conditions adopted during the simulations, all performed for $T=0.15$.}
\label{table1}
\begin{tabular*}{0.90\textwidth}{@{\extracolsep{\fill}}ccccccccc}
\hline
\hline
& $d$ & $\chi$ & $N_{\rm A}$ & $N_{\rm B}$ & $\rho$ & $\eta_{\rm A}$ & \# of runs\\
\hline
& 1 & 0.10 & 200 & 1800 & 0.05 & 0.002618 & 1\\
& 1 & 0.20 & 400 & 1600 & 0.05 & 0.005236 & 1\\
& 1 & 1/3 & 400 & 800 & 0.05 & 0.008727 & 1\\
& 1 & 0.50 & 400 & 400 & 0.05 & 0.013990 & 1\\
\hline
& 2 & 0.10 & 100 & 900 & 0.032 & 0.01340 & 1\\
& 2 & 0.20 & 200 & 800 & 0.016 & 0.01340 & 4\\
& 2 & 1/3 & 333 & 667 & 0.0096 & 0.01340 & 1\\
& 2 & 0.40 & 400 & 600 & 0.008 & 0.01340 & 1\\
& 2 & 0.50 & 400 & 400 & 0.0064 & 0.01340 & 1\\
\hline
& 3 & 0.10 & 100 & 900 & 0.01 & 0.01414 & 1\\
& 3 & 0.20 & 200 & 800 & 0.005 & 0.01414 & 4\\
& 3 & 0.20 & 400 & 1600 & 0.005 & 0.01414 & 4\\
& 3 & 0.20 & 400 & 1600 & 0.0025 & 0.00707 & 4\\
& 3 & 1/3 & 333 & 667 & 0.003 & 0.01414 & 1\\
& 3 & 0.40 & 400 & 600 & 0.0025 & 0.01414 & 1\\
& 3 & 0.50 & 400 & 400 & 0.002 & 0.01414 & 1\\
\hline
\hline
\end{tabular*}
\end{center}
\end{table}

We start each simulation run from a random configuration of the system to simulate its thermalization after a quench from high temperature (we have generated runs of 3-7 billion MC cycles). The relative amount of A and B is adjusted to the prescribed concentration $\chi$ of spheres; the temperature is $T=0.15$ and the density $\rho$ is at most $0.05$. We summarize in Table~\ref{table1} the conditions assumed in our runs. Four distinct runs have been performed for $d=2$ and $\chi=0.20$, as well as for each case relative to $d=3$ and $\chi=0.20$, so as to collect more statistics for the all-important case of self-assembly into crystalline membranes and vesicles. In the latter case, we have doubled the number of particles also to rule out any possible size-dependence of self-assembly results. The temperature of 0.15 is a compromise: sufficiently small to observe long-lasting aggregates, but still high enough to allow for escape from shallow energy minima.

In the production runs, which are typically $2\times 10^8$ cycles long, we compute various radial distribution functions (RDFs). Even in a strongly heterogeneous system, the sphere-sphere RDF $g_{\rm AA}(r)$ bears valuable information on the arrangement of spheres in a close neighborhood of a reference sphere. Useful indications on the relative separation of spheres and dimers are instead obtained from $g_{\rm AB_1}(r)$.

The fractal dimension of the subsystem of spheres can be obtained from $g_{\rm AA}(r)$: choosing a sphere as reference, for each fixed radius $R$ one counts the number of spheres within a distance $R$ from the reference sphere, finally averaging over all spheres. The outcome is the ``mass'' $M$ of the sphere backbone as a function of $R$. Typically, $M(R)\sim R^D$ at large $R$, which defines $D$ as the {\it mass fractal dimension} of the spheres. It follows immediately from the definition of the sphere-sphere RDF that
\be
M(R)=4\pi\rho\chi\int_0^R{\rm d}r\,r^2g_{\rm AA}(r)\,,
\label{eq:mass}
\ee
which allows one to readily obtain $D$ from $g_{\rm AA}(r)$.

To gain a better insight into the system structure we carry out a cluster analysis by identifying at regular times, and counting as a function of size, assemblies of connected spheres by the Hoshen-Kopelman algorithm~\cite{Hoshen}. Two spheres are connected if their distance is smaller than $r_{\rm min}=\sigma_{\rm A}+3\sigma_{\rm B_1}=(d+1)\sigma_{\rm B_2}$, representing the maximum distance at which two spheres can still be ``in contact'' through a B$_1$ monomer placed in the middle. Finally, the cluster-size distribution (over a fixed time interval) is defined as
\be
{\cal N}(n)=\frac{nN_{\rm cl}(n)}{N_A}\,,
\label{eq:CSD}
\ee
where $N_{\rm cl}(n)$ is the average number of $n$-sized clusters per system configuration. The distribution (S2) is so normalized that $\sum_n{\cal N}(n)=1$.

%
%
\begin{figure}[b!]
\begin{center}
\begin{tabular}{c}
\includegraphics[width=11cm]{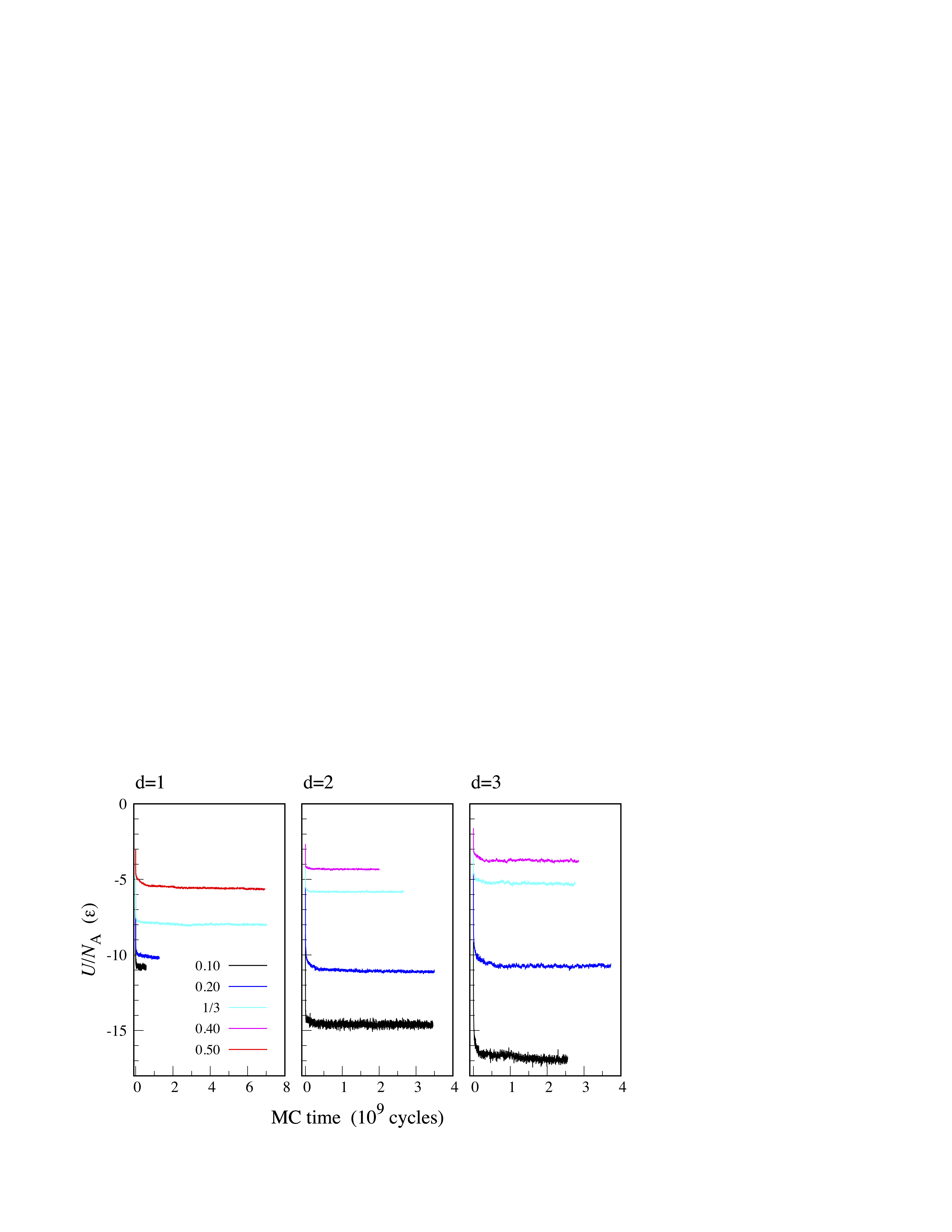} \\
\includegraphics[width=11cm]{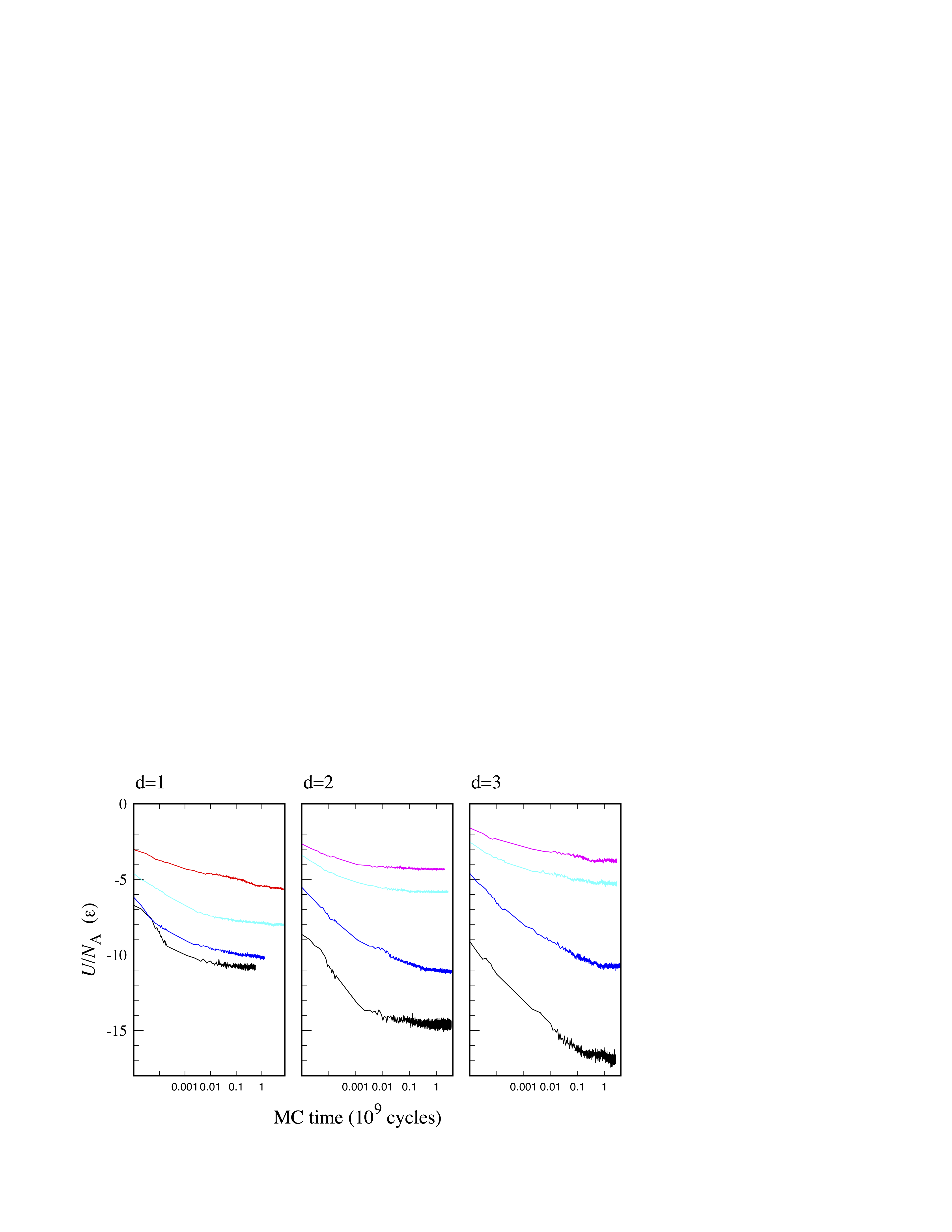}
\end{tabular}
\caption{MC evolution of the potential energy per sphere for a few $d$ and $\chi$ values (see legend). Top: standard linear scale. Bottom: semi-log scale.}
\label{fig:energy}
\end{center}
\end{figure}

\section{RESULTS AND DISCUSSION}

Simulations are kept going until the total potential energy $U$ fluctuates around a constant value for long (a few billion cycles), signaling that a (meta)stable equilibrium has been reached at last. In Fig.~\ref{fig:energy}, we report some representative cases of $U$ evolution in the course of simulation. Each curve refers to an individual run, i.e. no average is made over several runs performed under identical conditions. As evidenced by a glance at the configuration of the system at regular intervals, relaxation initially proceeds through the progressive accretion of a few sphere aggregates, glued together by dimers, which is reflected in a rapid (exponential) decay of $U$. Subsequently, $U$ decreases more slowly as clusters begin to coalesce, until it levels off after $\approx 10^9$ cycles. A slowing down of relaxation (aging) will also occur for large $\chi$ ($>0.50$), where the formation of extended sphere aggregates is hampered by the shortage of dimers. Any coagulation event is manifested in a tiny downward jump of $U$ as a function of time; however, since the joining of clusters is a relatively rare event, at least on the timescale of single-particle diffusion, the decrease of $U$ is slower in the late stage of the evolution than in the first stages. On the other hand, a well-definite drop of $U$ as a function of MC time signals an extensive rearrangement of the structure, as clearly seen for $d=3$ and $\chi=0.10$ (black lines in the right panels of Fig.~\ref{fig:energy}).

\begin{figure}[t]
\begin{center}
\begin{tabular}{c}
\includegraphics[width=11cm]{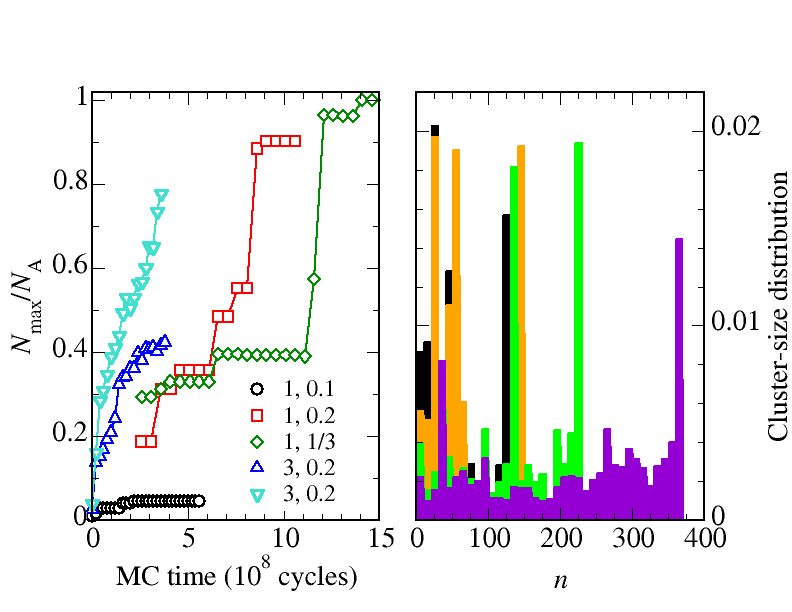}
\end{tabular}
\caption{Left: Size $N_{\rm max}$ of the maximum cluster of spheres computed at discrete times along the MC trajectory for $N_{\rm A}=400$ and a few combinations of $d$ and $\chi$ (see legend). The data for $d=1$ refer to $\rho=0.05$; the data for $d=3$ are averages over four distinct runs (upward triangles: $\rho=0.0025$; downward triangles: $\rho=0.005$). Right: Cluster-size distribution for $d=1$ and $\chi=0.20$, as computed over four distinct intervals of $10^8$ cycles separated by $2\times10^8$ cycles (temporal sequence: black, orange, green, purple).}
\label{fig:clustersize}
\end{center}
\end{figure}

As aggregates grow in size and relax, the dimers on the surface become increasingly effective in screening the aggregate from the outside particles. Indeed, a growing aggregate gets progressively covered with the inert B$_2$ particles, while the ``reactive'' particles (i.e. B$_1$ monomers and spheres) lie buried within. The next step in equilibration is cluster coalescence, aka coarsening/Ostwald ripening, which for $T=0.15$ typically starts $10^7$-$10^8$ cycles after the initial quench. When two clusters meet, they usually stick together to form a bigger aggregate. The kinetics of coarsening is strongly influenced by the system density, i.e. by the crowding around the aggregates, which affects the rate of collisions and indirectly the type of frozen architectures arising at low temperature. Unless the initial concentration of spheres is very low, a unique aggregate encompassing all spheres in the system eventually appears, as witnessed by the MC evolution of the maximum-cluster size in the left panel of Fig.~\ref{fig:clustersize}. Details are provided in the right panel, where we show how the cluster-size distribution evolves in a single run for $d=1$ and $\chi=0.20$: as coarsening proceeds, the weight of small sizes is progressively reduced in favor of the maximum-cluster size, until a unique peak centered at the total number $N_{\rm A}$ of spheres (400 in this figure) is left over.

%
%

\begin{figure}[t!]
\begin{center}
\begin{tabular}{c}
\includegraphics[width=16cm]{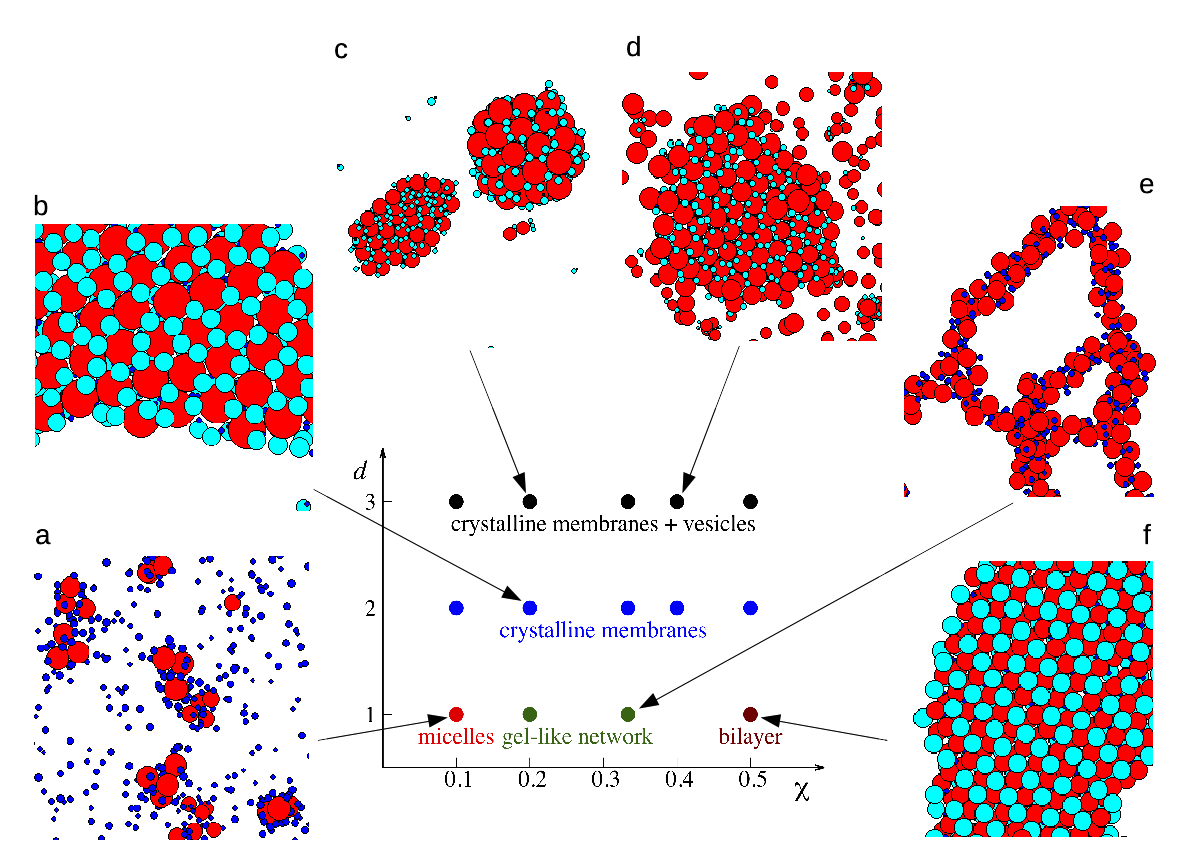}
\end{tabular}
\caption{Schematic diagram of self-assembly at low density (circles denote the state points where simulations are carried out). For selected $(\chi,d)$ pairs, we show the typical structure of the system at equilibrium (spheres are in red; B$_1$ and B$_2$ monomers are in blue and cyan, respectively). Only a portion of the system is depicted in the snapshots. In (a) and (e) B$_2$ particles are not shown for clarity.}
\label{fig:phasediagram}
\end{center}
\end{figure}

We sum up our findings in the $(\chi,d)$ diagram shown in Fig.~\ref{fig:phasediagram}, yielding a bird's eye view of the self-assembled structures developed in the system at low density and temperature. In the following, we separately discuss results for $d=1$ and $d=2,3$ in the subsections~\ref{sec:3.1}
and~\ref{sec:3.2}, respectively. Finally, in subsection~\ref{sec:3.3} 
we show by analytic arguments that the nature of self-assembly for $d=1$ is necessarily different from $d=2$ or 3.

%
%

\subsection{d=1} \label{sec:3.1}

\begin{figure}
\begin{tabular}{ccc}
\includegraphics[width=5.0cm]{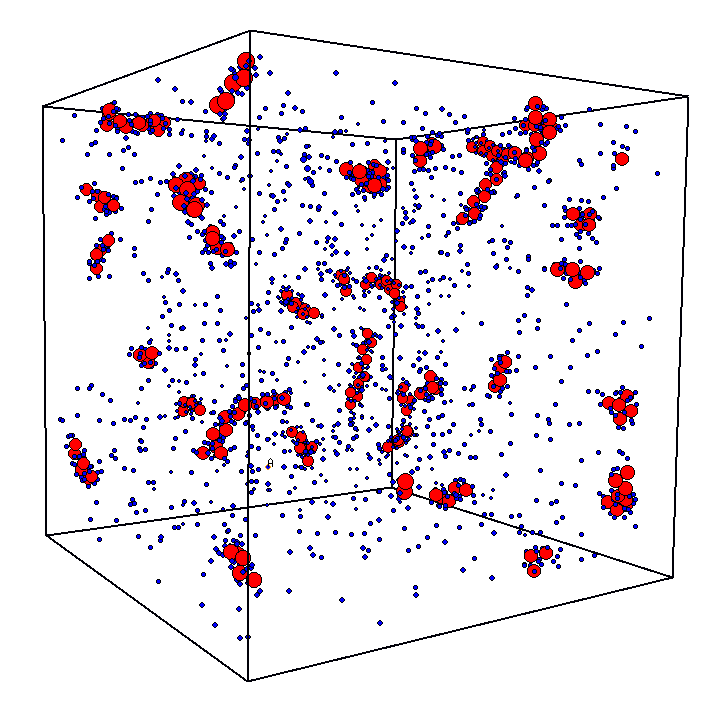} &
\includegraphics[width=5.0cm]{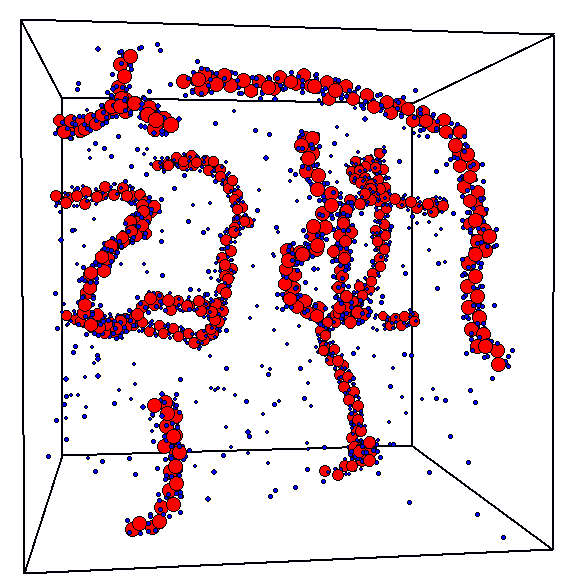} &
\includegraphics[width=5.0cm]{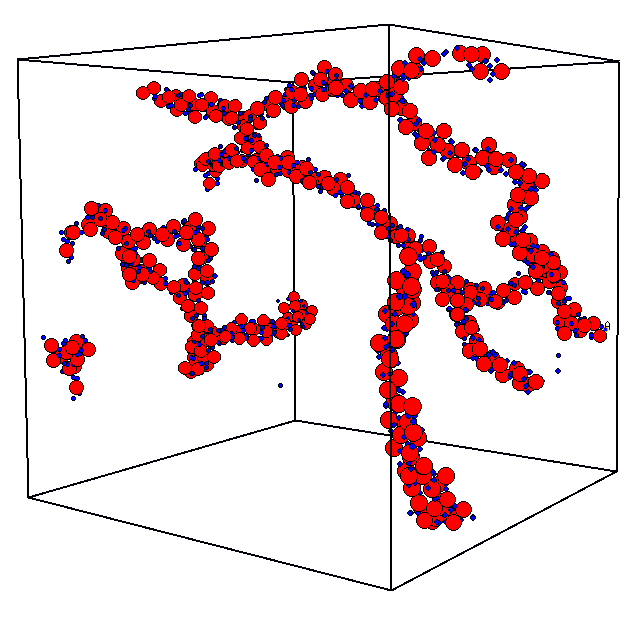}
\end{tabular}
\caption{Aggregates for $d=1$: micelles ($\chi=0.10$, left) and gel-like networks ($\chi=0.20$, middle; $\chi=1/3$, right). B$_2$ particles are hidden for a better visualization of the sphere backbone.}
\label{fig:d1}
\end{figure}

For $d=1$ we set the density at $0.05$ and consider $\chi$ values from 0.10 to 0.50. When $\chi=0.10$ we observe a homogeneous distribution of small clusters of spheres coated with dimers (``micelles'', see Fig.~\ref{fig:phasediagram}a). We see in the left panel of Fig.~\ref{fig:d1} that both spherical and rod-like micelles are formed. The dimers around the spheres, exposing B$_2$ monomers outside, are so tight together to prevent coalescence of clusters. Therefore, aggregation of spheres is successfully contrasted. In a similar way, stabilization of polystyrene microspheres by dumbbell-shaped colloids with a sticky and a non-sticky lobe was demonstrated in Ref.~\citenum{Wolters}.

%
%
 
%
%
\begin{figure}[!b]
\begin{center}
\begin{tabular}{cc} 
\includegraphics[width=5cm]{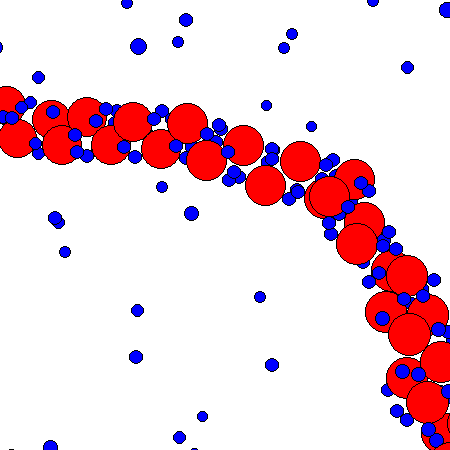} \hspace{2cm} &
\includegraphics[width=5cm]{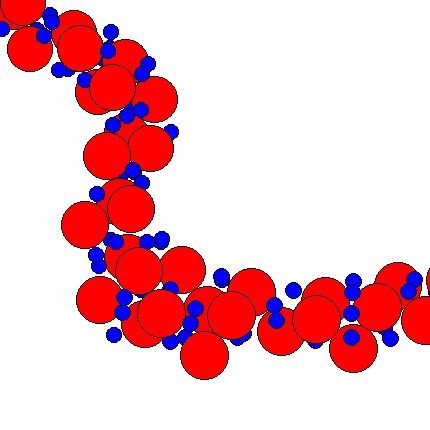}
\end{tabular}
\end{center}
\caption{Chain details of gel-like networks for $d=1$: $\chi=0.20$, left; $\chi=1/3$, right. B$_2$ particles are not shown.}
\label{fig:chains}
\end{figure}

Clearly, micellization is only stable provided that the concentration $\chi$ is low enough, and this is apparently the case for $\chi=0.10$. For larger concentrations, spheres eventually become part of a unique aggregate, whose nature depends on $\chi$. For $\chi=0.20$ and $\chi=1/3$ a percolating network is formed (see Fig.~\ref{fig:phasediagram}e, and the middle and right panels of Fig.~\ref{fig:d1}). Indeed, spheres are now numerous enough to rule out the occurrence of spherical micelles, opening the way to the formation of long chains of spheres. Each sphere along the chain binds from eight to ten dimers (depending on $\chi$), to be shared with its neighboring spheres, as witnessed by the energy level in Fig.~\ref{fig:energy} (left panels). Indeed, the absolute value of $U/N_{\rm A}$ is the mean number of B$_1$ monomers bound to a sphere. A close scrutiny of the chains for $\chi=0.20$ reveals a zig-zag structure (Fig.~\ref{fig:chains}, left), a motif also present in rod-like micelles for $\chi=0.10$. For $\chi=1/3$ the chain geometry with the most effective combination of low energy and large entropy is given by three helicoidal strands of spheres wrapped around a common (curved) axis (Fig.~\ref{fig:chains}, right). During the MC evolution, the presence of unsaturated bonds on the chain surface makes crossing/branching of chains a likely outcome, eventually resulting in a connected network that percolates throughout the simulation box. Once this spanning network has been established (which takes $\approx 10^9$ MC cycles for $T=0.15$), no further rearrangement occurs at the large scale: the aggregate has become a gel. We have computed the mass fractal dimension of the backbone of connected spheres and found it to be close to 1.8, seemingly at odds with the intuition that a chain network is a one-dimensional manifold. In fact, a fractal dimension between 1 and 2 is a reasonable outcome, considering that chains have a non-vanishing thickness and, due to the high number of nodes in the network, a relevant fraction of spheres are not located in the body of a chain.

%
%
\begin{figure}[!b]
\begin{tabular}{c}
\includegraphics[width=10cm]{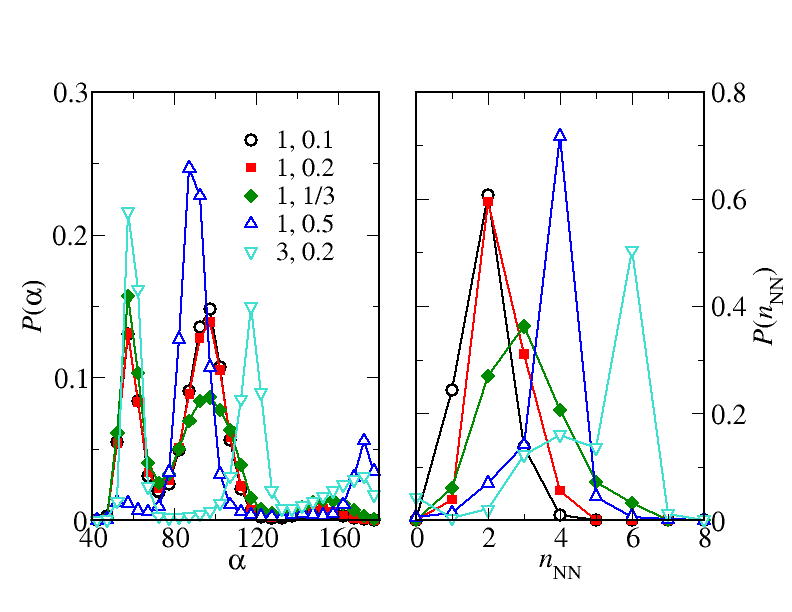}
\end{tabular}
\caption{Left: Distribution of the angle $\alpha$ formed by two A-A bonds sharing one sphere at the angle vertex, for a few combinations of $d$ and $\chi$ (see legend). The data for $d=1$ and $d=3$ refer to $\rho=0.05$ and $0.0025$, respectively. Right: Distribution of the number $n_{\rm NN}$ of spheres that are nearest neighbors to a given sphere.}
\label{fig:angles}
\end{figure}

Looking at the distribution of the angle $\alpha$ between two A-A bonds with one sphere in common at the angle vertex, we see a resemblance between $\chi=0.10$ and $\chi=0.20$ (compare circles and squares in the left panel of Fig.~\ref{fig:angles}), to be ascribed to the similarity in the local structure between rod-like micelles and gels. In detail, while the peak centered at 60$^\circ$ points to the existence of a high number of triplets of spheres in reciprocal contact, the broader peak around $100^\circ$ indicates a preference for local cubic and tetrahedral orderings, as well as for zig-zag order with this characteristic angle. Upon going from $\chi=0.10$ to $\chi=1/3$, the mean number of spheres that are nearest neighbors to a reference sphere, $n_{\rm NN}$, grows from $\lesssim 2$ to $\gtrsim 3$ (Fig.~\ref{fig:angles}, right).

%
%
\begin{figure}[t]
\begin{tabular}{cc}
\includegraphics[width=6cm]{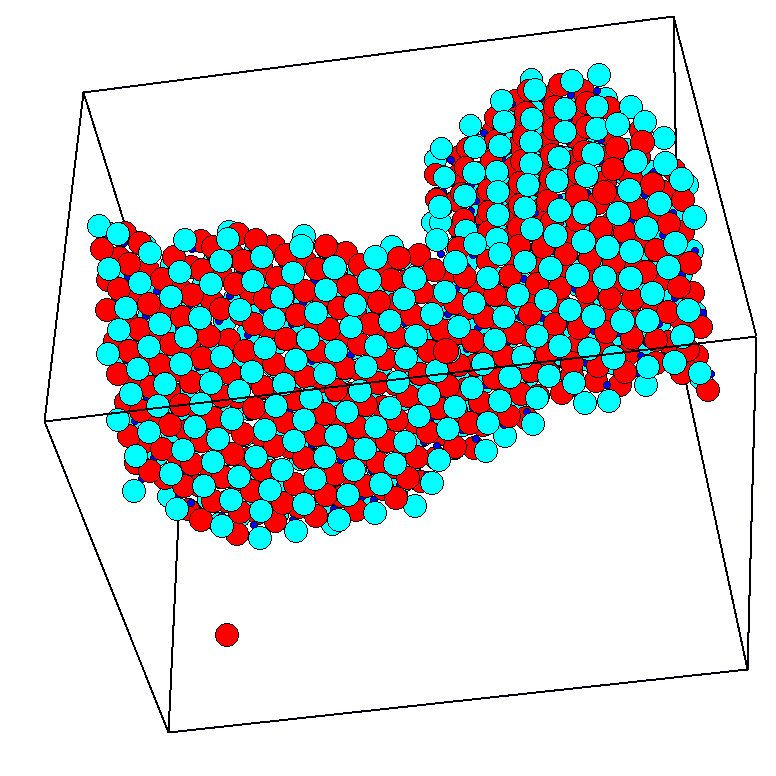} &
\includegraphics[width=6cm]{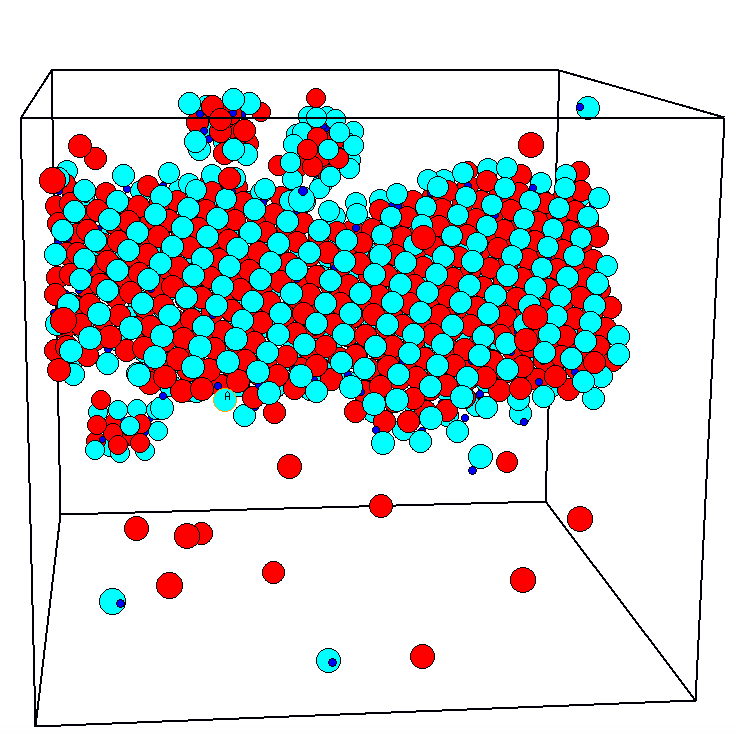}
\end{tabular}
\caption{Crystalline bilayer ($d=1,\chi=0.50$). Left: $T=0.15$. Right: $T=0.20$.}
\label{fig:bilayer}
\end{figure}

With a further increase in concentration the geometry of self-assembly changes again: for $\chi=0.50$ we see the onset of a crystalline bilayer, growing laterally through the inclusion of spheres and dimers from the solution (Figs.~\ref{fig:phasediagram}f and \ref{fig:bilayer}). The arrangement of particles in the bilayer is very peculiar: each layer of spheres forms a rectangular crystal displaced by half lattice spacing relative to the other layer; B$_1$ particles occupy the interstices between the spheres, so that each sphere is bound to exactly six dimers (see details in subsection~\ref{sec:3.3}). This structure is reflected in the $\alpha$ distribution (Fig.~\ref{fig:angles}, left), showing a distinct peak at 90$^\circ$, and in the $n_{\rm NN}$ distribution, peaked at 4 (Fig.~\ref{fig:angles}, right). The mass fractal dimension is now about 2.1, consistent with the idea that a bilayer is essentially two-dimensional. When heating the system from $T=0.15$ to $T=0.20$, we find that the bilayer structure keeps stable, even though small clusters of spheres and dimers detach from the bilayer edge to reach the solution (Fig.~\ref{fig:bilayer}, right).

%
%
\begin{figure}[!t]
\begin{tabular}{c}
\includegraphics[width=8cm]{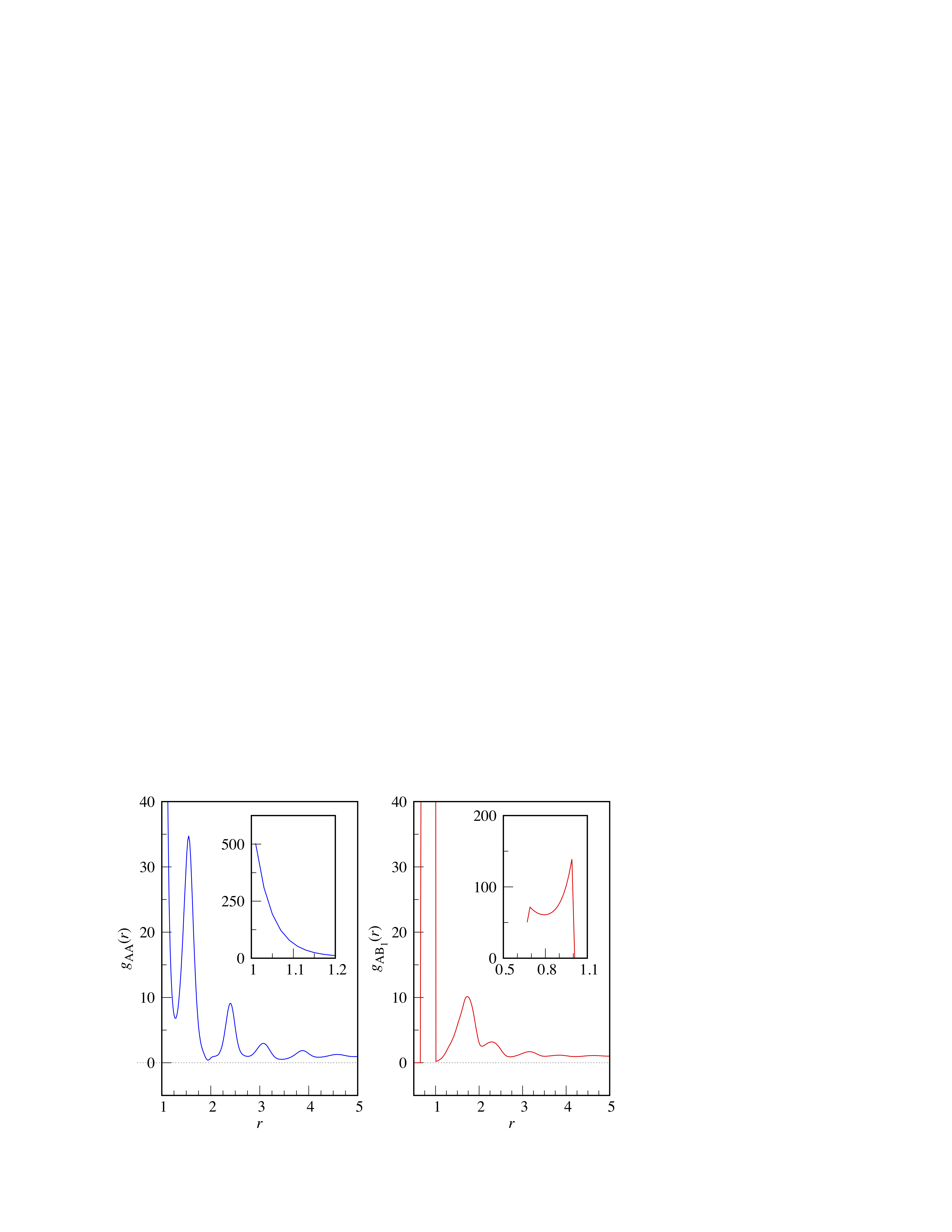} \\
\includegraphics[width=8cm]{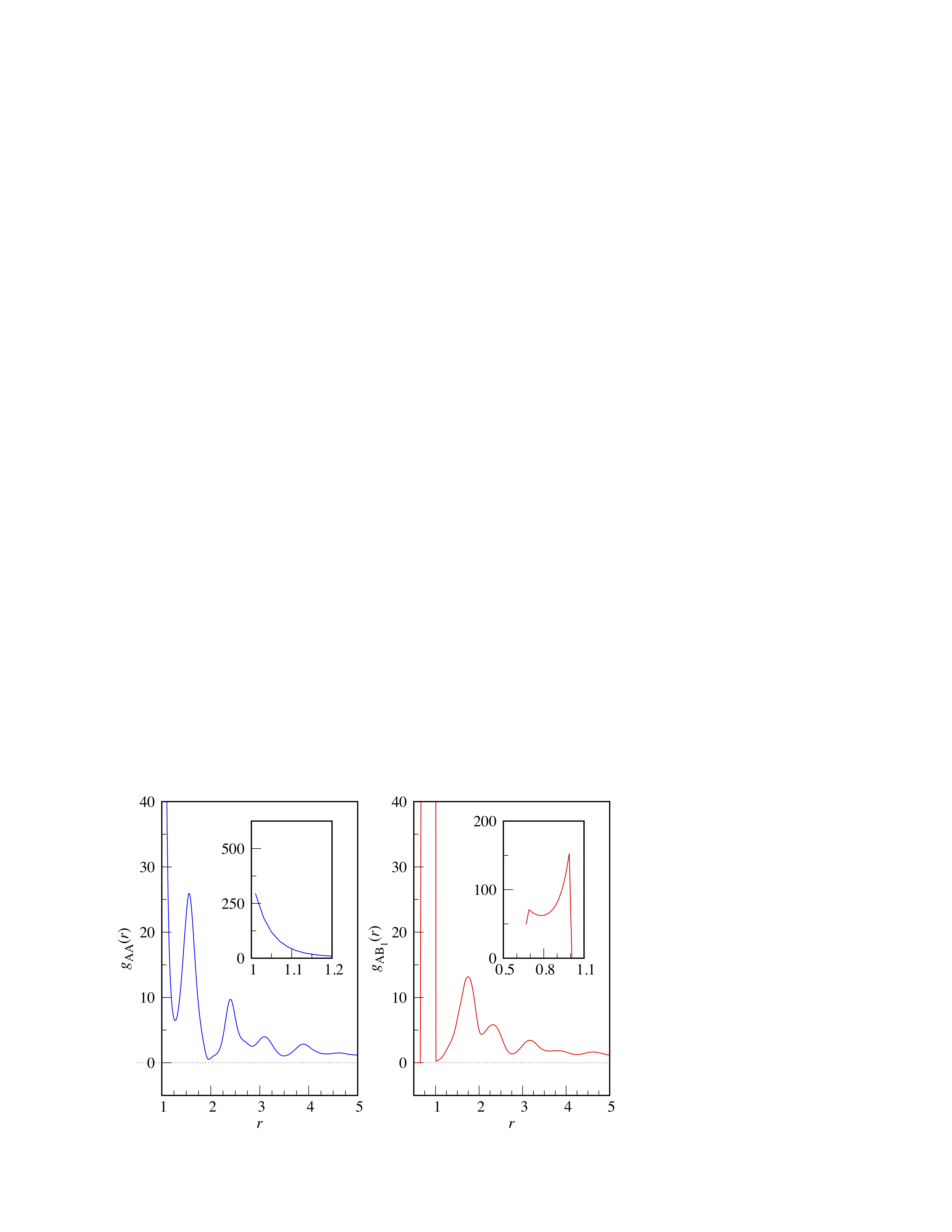} \\
\includegraphics[width=8cm]{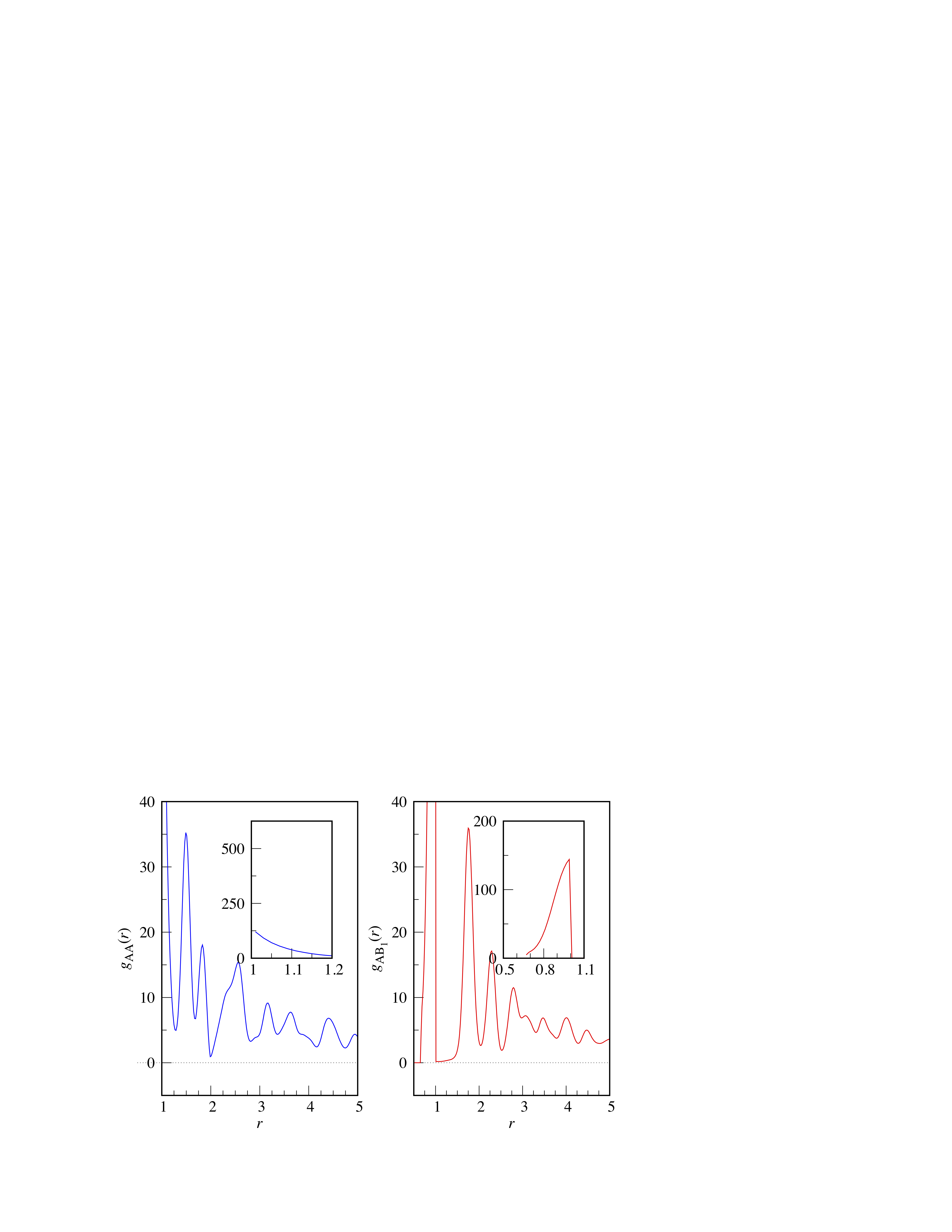}
\end{tabular}
\caption{RDF $g_{\rm AA}(r)$ (left) and $g_{\rm AB_1}(r)$ (right) for $d=1$ and $\rho=0.05$. From top to bottom, curves refer to $\chi=0.10$, 0.20 and 0.50. A magnification of the contact region is reported in the insets.}
\label{fig:RDF1}
\end{figure}

The spatial distribution of spheres at a local level has also been investigated by RDFs (see Fig.~\ref{fig:RDF1}), confirming the structural similarity between $\chi=0.10$ and $\chi=0.20$. Upon increasing the concentration, the maximum of $g_{\rm AA}(r)$ at contact is progressively reduced as the system departs more and more from micelles. In the range $\chi=0.10$\,-$1/3$ the second peak is associated with bound pairs of spheres separated by a B$_1$ particle. We see that it broadens until, for $\chi=0.50$, it splits in two (we comment more on this point in subsection~\ref{sec:3.3}).

\subsection{d=2, 3} \label{sec:3.2}
The value of $d$ is rather crucial for the stability of the crystalline bilayer, which no longer exists for $d$ larger than $\approx 1.35$ (see the theoretical argument in subsection~\ref{sec:3.3}). Indeed, when $d=2$ or 3 the nature of self-organized structures is different. Aggregates are now monolayer sheets in the whole $\chi$ range from 0.10 to 0.50, made up of sixfold-coordinated spheres held together by dimers (``crystalline membranes'', often resulting from the fusion of smaller patches), see Figs.~\ref{fig:phasediagram}b-d: each sphere is bound to twelve dimers on average (six on each face of the sheet), located in the interstices between triplets of neighboring spheres. The finite range of attraction allows for some tolerance in the separation between spheres and in the position of intercalated dimers; as a result, membranes are not perfectly flat but can bend to a certain extent. Crystalline membranes usually coexist with a sizable number of isolated dimers ($\chi<0.20$) or spheres ($\chi>0.20$). Only for $\chi=0.20$ the right proportion of spheres and dimers is reached for building lamellar aggregates without excess particles. In subsection~\ref{sec:3.3} we provide a theoretical argument ruling out the existence of membranes for $d=1$.

%
%
\begin{figure}
\begin{center}
\begin{tabular}{ccc}
\includegraphics[width=5cm]{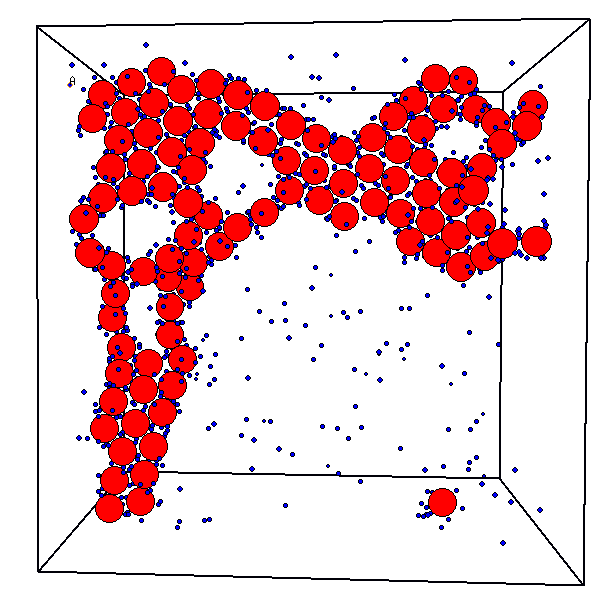} \ \ \ &
\includegraphics[width=5cm]{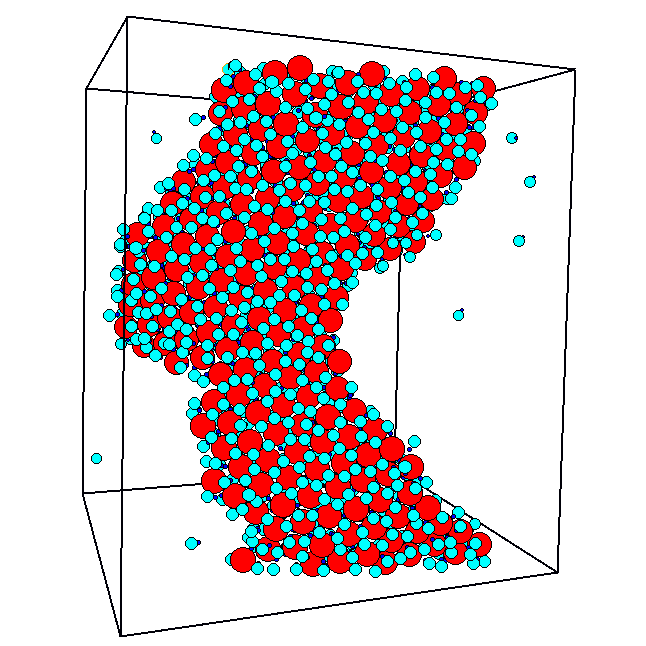} \ \  \ &
\includegraphics[width=5cm]{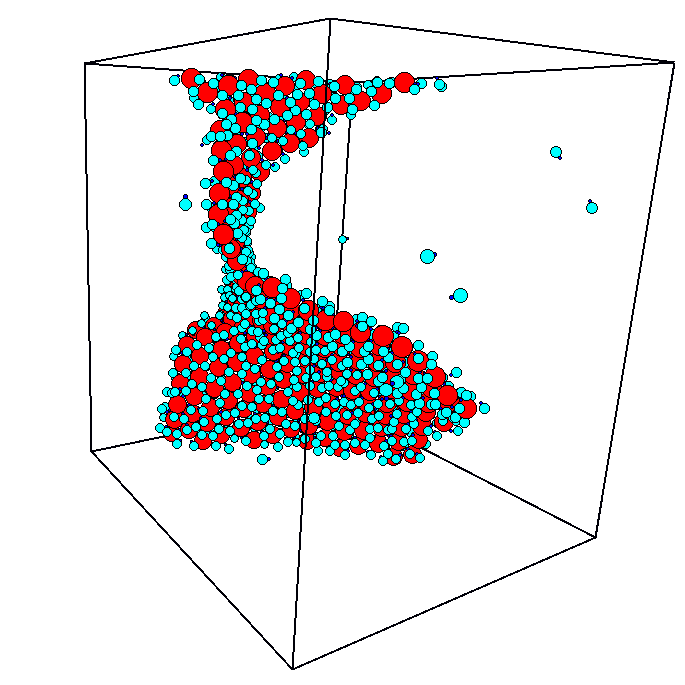}
\end{tabular}
\caption{Left: membrane with holes ($d=2$ and $\chi=0.10$; B$_2$ particles are not shown). Middle: A more conventional membrane for $d=2,\chi=0.20$, and $\rho=0.016$. Right: M\"{o}bius strip ($\chi=0.20$ and $\rho=0.016$).}
\label{fig:oddities}
\end{center}
\end{figure}

The rich catalog of membrane morphologies can be appreciated in Fig.~\ref{fig:oddities}, which is relative to $d=2$. For $\chi=0.10$ the observed aggregates are one-layer sheets with holes (left). When increasing the concentration to 0.20, more conventional membranes are observed (middle), which are flat or only slightly curved. Occasionally, more exotic structures are seen: an example is shown in the right panel of Fig.~\ref{fig:oddities}, reporting a twisted crystalline membrane which gives an atomistic representation of a M\"{o}bius strip.

%
%
\begin{figure}[!b]
\begin{tabular}{c}
\includegraphics[width=11cm]{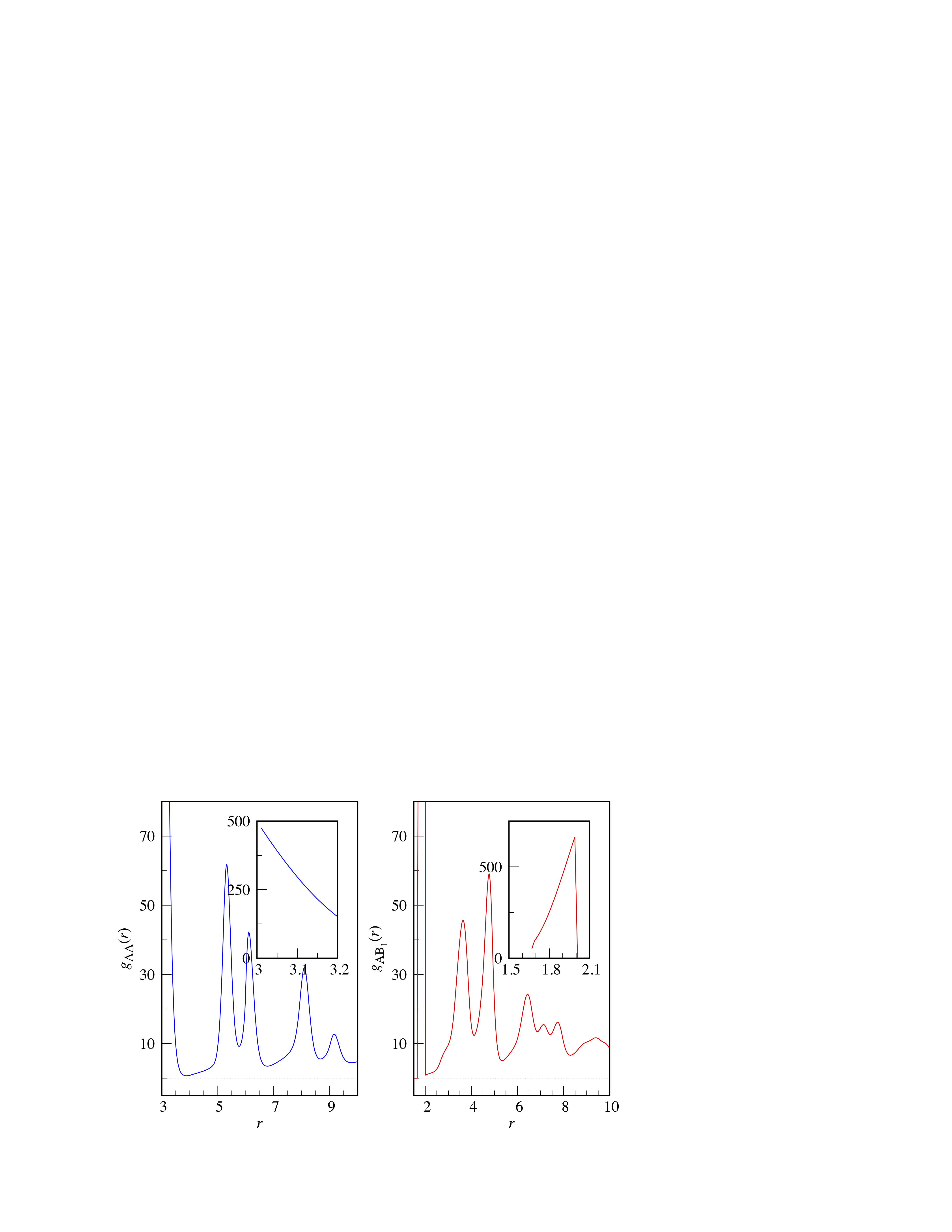}
\end{tabular}
\caption{RDF $g_{\rm AA}(r)$ and $g_{\rm AB_1}(r)$ for $d=3,\chi=0.20$, and $\rho=0.0025$. A magnification of the contact region is reported in the insets.}
\label{fig:RDF3}
\end{figure}

The triangular array of spheres within membranes has clear imprints in the structural indicators. This is evidenced in the narrow peak at 60$^\circ$ in the $\alpha$ distribution (Fig.~\ref{fig:angles}, left), with replicas at 120$^\circ$ and 180$^\circ$, as well as in the maximum at 6 in the $n_{\rm NN}$ distribution (Fig.~\ref{fig:angles}, right). The triangular order is also revealed in the locations of the first few peaks in the sphere-sphere RDF (Fig.~\ref{fig:RDF3}): indeed, for $d=3$ and $\chi=0.20$ we see that the first three shell radii of the triangular lattice all occur in the profile of $g_{\rm AA}(r)$, whereas the first peak of $g_{\rm AB_1}(r)$ corresponds to B$_1$ particles located in the interstices between triplets of neighboring spheres. 

Looking back at Fig.~\ref{fig:energy}, we can now explain that the different $d$ dependence of the asymptotic value of $U$ for fixed $\chi$ is due to the diversity of self-assembly solutions devised by the system as a function of $d$. For $\chi=0.10$ there is no preferential site for binding dimers to a sphere; therefore, the asymptotic value of $|U|/N_{\rm A}$ is an increasing function of $d$ for the simple reason that a bigger sphere can bind a larger number of dimers. On the contrary, for $\chi=1/3$ or 0.50 the decrease of $|U|/N_{\rm A}$ with $d$ just demonstrates the higher efficiency of spheres to bind dimers for $d=1$, where a gel-like network ($|U|/N_{\rm A}=8$-10) or a crystalline bilayer ($|U|/N_{\rm A}=6$) occur. For $d=2$ or $3$, one might expect that $U/N_{\rm A}\simeq -12$ (since the specific energy in a crystalline membrane); however, for $\chi>0.20$ membranes are floating in a background of unbound spheres, and this pushes the asymptotic value of $U/N_{\rm A}$ upwards relative to $-12$, the more so the larger $\chi$ is relative to 0.20.

\begin{figure}[t]
\begin{tabular}{ccc}
\includegraphics[width=4.8cm]{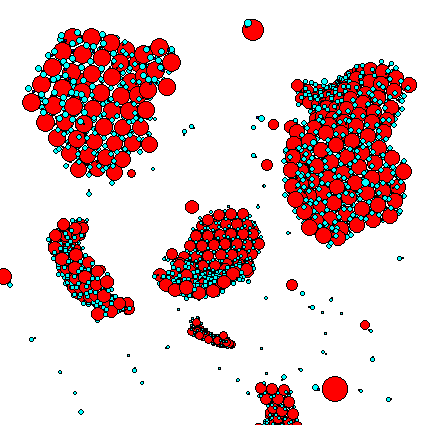} \ \ \ &
\includegraphics[width=4.8cm]{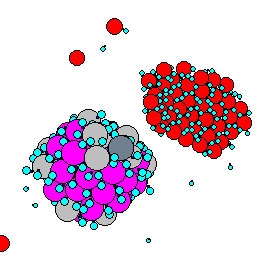} \ \ \  &
\includegraphics[width=4.8cm]{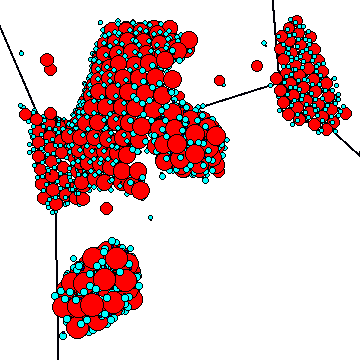}
\end{tabular}
\caption{Left: a typical configuration of the A-B mixture for $d=3,\chi=0.20$, and $\rho=0.0025$. Middle: vesicle with a membrane in the background; in the vesicle, 4-fold, 5-fold, and 6-fold spheres (46 in total) are colored in slate grey, light grey, and magenta, respectively. Right: lamellar aggregates for $d=3,\chi=0.20$, and $\rho=0.005$.}
\label{fig:membranes}
\end{figure}

The spontaneous generation of flexible membranes is an attractive feature of our mixture, since it brings about the possibility of vesicle formation from scratch (at least under high-dilution conditions). Vesicles have long been recognized as a fundamental requisite of life,~\cite{Chen} since all known life forms are cellular and each cell is screened from the environment by a closed bilayer shell composed by lipids and proteins in a fluid state. Molecular simulations~\cite{Bernardes,Noguchi3,Yamamoto,Cooke,Lenz,Noguchi} and small-angle X-ray scattering experiments~\cite{Weiss} have shown that vesiculation of amphiphiles usually proceeds from small discoidal patches which, beyond a certain size, are energetically preferred to spheroidal grains. Gradually, these patches grow by joining individual amphiphiles to the peripheral contour. To reduce contour energy, a sufficiently large patch acquires a definite curvature, until it eventually folds into a vesicle.~\cite{Fromherz,Noguchi2} Notably, in our model the onset of vesicles follows the same path, but for the difference that membrane sheets are now crystalline and one-layer (see Fig.~\ref{fig:membranes}, left). Clearly, the decrease in contour energy during the transition from a planar sheet to a vesicle is hampered by an increase in bending energy due to stretching of bonds, leading ultimately to the existence of a minimum membrane radius below which the formation of vesicles is unfavorable.~\cite{Israelachvili,Huang} Crystalline membranes can circumvent this limitation by the proper insertion of a few fivefold disclinations, relieving the strain associated with the defect by buckling out of the plane.~\cite{Nelson,Seung,Lidmar,Kohyama,Haselwandter,Bowick} The same mechanism is at work in our model, where the appearance of disclinations in suitable locations makes a small membrane sheet capable to transform into a vesicle (Fig.~\ref{fig:membranes}, middle). Around each conical protrusion several facets merge together to form a cup-shaped intermediate stage in vesicle development. A similar process occurs in small spherical crystals of hard particles,~\cite{Prestipino2,Prestipino3,Wales,Guerra} where the gathering of disclinations at the vertices of an icosahedron (i.e. at the maximum possible relative distance) guarantees the largest possible entropy, i.e. the highest number of sixfold particles. For $d=2$, membrane sheets appear to be flatter, i.e. stiffer as compared to $d=3$. This is due to a stronger mutual obstruction of nearby B$_2$ monomers in curved membranes (see details in subsection~\ref{sec:3.3}); as a result, in small system samples the propensity of membranes to evolve into vesicles for $d=2$ is smaller than for $d=3$.

%
%

We have already mentioned the importance of vesicles as ``containers'' of molecules of fundamental importance for life. This strongly depends on the lipid vesicle being impenetrable to most solutes. Interestingly, owing to the occupation of sphere interstices by dimers, also our colloidal vesicles are impermeable, making encapsulation effective.

It seems inevitable for any sufficiently flexible membrane sheet in our model to eventually form a vesicle. However, during the simulation run two aggregates may occasionally collide and join together. In particular, a forming vesicle may encounter another aggregate on its path and then vesiculation gets arrested. Clearly, a vesicle can only arise as an individual entity if the characteristic time for a membrane to fold and close is shorter than the average collision time. In this regard, in the right panel of Fig.~\ref{fig:membranes} we note the simultaneous occurrence of two vesicles, one of which is bound to a curved membrane. While the presence of two vesicles in a small sample is to be regarded as an exceptional event, it is confirmed that in our system the natural tendency of membrane sheets is folding into vesicles.

\subsection{Stability of Crystalline Structures} \label{sec:3.3}
In this subsection we prove that ({\it i}) the crystalline bilayer found for $d=1$ and $\chi=0.50$ cannot survive above $d\approx 1.35$; conversely, ({\it ii}) a triangularly-ordered sheet (i.e. a flat membrane of infinite size) cannot exist for $d$ smaller than $\approx 1.5$.

{\em Bilayer} --- Looking closely at the configuration pictured in the left panel of Fig.~\ref{fig:bilayer}, relative to $d=1$ and $\chi=0.50$, we see that spheres are arranged in two rectangular layers with lattice spacings $a$ and $b$ (with $a>b$), displaced by $c$ in the third direction. Dimers are oriented perpendicularly to the layers, such that the B$_1$ monomers are nested in the interstices of each layer, whereas the B$_2$ monomers are exposed to the outside. In a convenient reference system, the coordinates of spheres in the bottom layer are $(ai,bj,0)$ with $i,j=0,\pm 1,\pm 2,\ldots$, while those of spheres in the top layer are $(ai+a/2,bj,c)$. In turn, the B$_1$ particles in the bottom layer have coordinates $(a/2+ai,b/2+bj,0)$, whereas those in the top layer are located at $(ai,b/2+bj,c)$. Hence, the four B$_1$ particles that are in-plane neighbors of a given sphere are at a distance of $d_{\rm in}=\sqrt{a^2+b^2}/2$, while the two B$_1$ particles that are out-of-plane neighbors of a sphere are $d_{\rm out}=\sqrt{b^2+4c^2}/2$ far apart. Instead, the distance between two closest spheres lying in different layers is $d_{\rm ss}=\sqrt{a^2+4c^2}/2$. For symmetry reasons it must be $d_{\rm in}=d_{\rm out}$, which implies $c=a/2$ and $d_{\rm ss}=a/\sqrt{2}$. Both $d_{\rm ss}$ and $b$ should be larger than $\sigma_{\rm A}=1$, implying $a\ge\sqrt{2}$, $b\ge 1$, and $c\ge\sqrt{2}/2$. When all equal signs apply, $d_{\rm in}=d_{\rm out}=\sqrt{3}/2$. In order to have a bond between a sphere and a dimer, the relative distance between A and B$_1$ must satisfy the condition $\sigma_{\rm AB_1}\le r\le\sigma_{\rm AB_1}+\sigma_{\rm B_1}$, that is $2/3\le r\le 1$, which indeed is met by $r=\sqrt{3}/2$.

Clearly, thermal fluctuations impose an effective nearest-neighbor distance ($b$) which is not exactly 1. We may ask how much $a$ and $b$ can be scaled up relative to their limiting values, so that we still have six A-B$_1$ bonds per sphere. Hence we take, for a generic value of $\sigma_{\rm A}$, $a=\sqrt{2}(1+\delta)\sigma_{\rm A},b=(1+\delta)\sigma_{\rm A}$, and  $c=a/2$, where $\delta>0$ is the relative increment of distances due to fluctuations. Binding is guaranteed if $\sigma_{\rm AB_1}\le\sqrt{a^2+b^2}/2\le\sigma_{\rm AB_1}+1/3$, or
\be
\sigma_{\rm min}\equiv\frac{1}{3\sqrt{3}-3+\delta}\le\sigma_{\rm A}\le\sigma_{\rm max}\equiv\frac{1}{\sqrt{3}(1+\delta)-1}\,.
\label{eq:bilayer}
\ee
While $\sigma_{\rm min}<1$, $\sigma_{\rm max}$ is a decreasing function of $\delta$, assuming a value of $1.366$ for $\delta=0$. Hence we conclude that no crystalline bilayer of the kind observed for $d=1$ can exist for $d=2$ or 3. Looking at $g_{\rm AA}(r)$ for $\sigma_{\rm A}=1$ (Fig.~\ref{fig:RDF1}, bottom), we see a first-neighbor peak at 1 (corresponding to the minimum value of $b$ or, equivalently, $d_{\rm ss}$), a second-neighbor maximum at $\simeq 1.49$ (corresponding to $a$) and a third-neighbor maximum at $\simeq 1.83$ (corresponding to $\sqrt{a^2+b^2}$). From the value of $a$ we get $\delta=0.054$, which gives $b=1.054$ and $\sqrt{a^2+b^2}=1.825$, as indeed observed.

\begin{figure}[t]
\begin{tabular}{ccc}
\includegraphics[width=5.2cm]{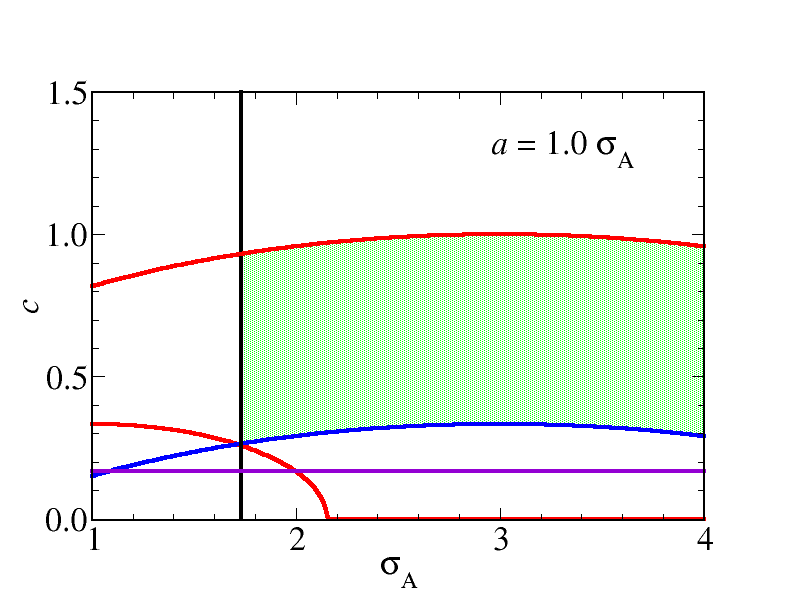} &
\includegraphics[width=5.2cm]{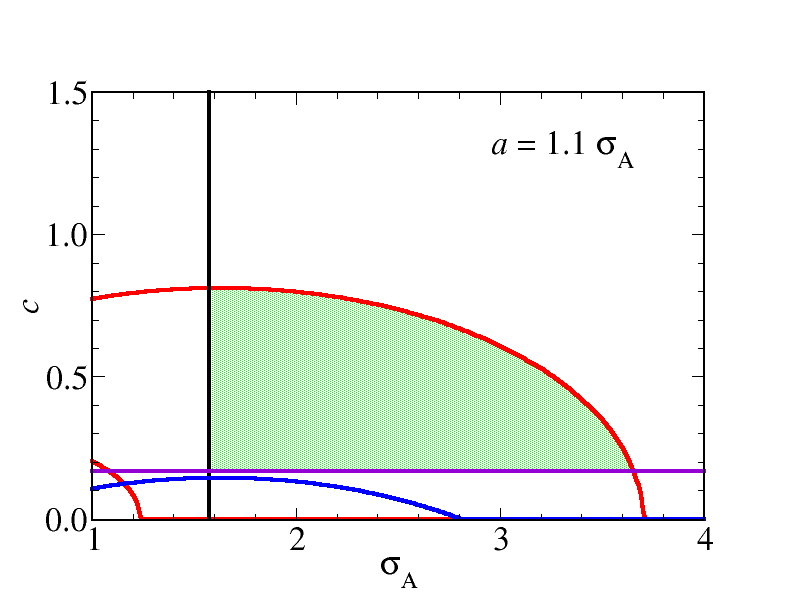} &
\includegraphics[width=5.2cm]{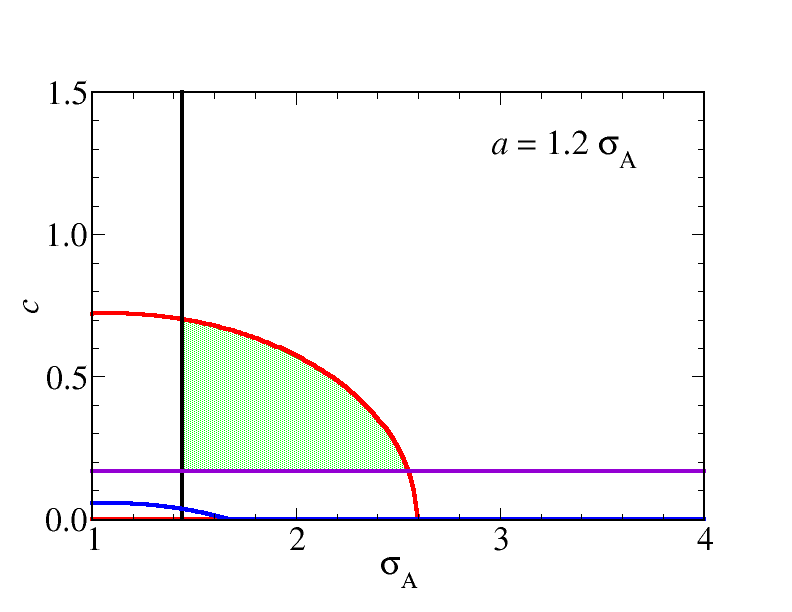}
\end{tabular}
\caption{Graphical solution of Eqs.\,(\ref{eq:sheet1})-(\ref{eq:sheet3}). For each $a$ considered, the allowed values of $\sigma_{\rm A}$ and $c$ fall within the green region. Both $\sigma_{\rm A}$ and $c$ are given in units of $\sigma_{\rm B_2}$.}
\label{fig:theory}
\end{figure}

{\em Triangular sheet} --- Let us consider a triangular carpet of spheres, with spacing $a\ge\sigma_{\rm A}$. Dimers are located in correspondence of the interstices of the sphere crystal, on both sides of it, and are perpendicularly oriented. In a convenient reference frame, the coordinates of three neighboring spheres are $(0,0,0),(a,0,0)$, and $(a/2,a\sqrt{3}/2,0)$. Calling $c$ the distance of a B$_1$ monomer from the plane of spheres, the coordinates of the B$_1$ and B$_2$ particles which are equidistant from the three spheres are, on the upper side of the plane, $(a/2,a\sqrt{3}/6,c)$ and $(a/2,a\sqrt{3}/6,c+2/3)$. Taking into account the B$_1$ monomer that is placed symmetrically below the plane of spheres, the smallest $c$ allowed is $\sigma_{\rm B_1}/2=\sigma_{\rm B_2}/6$. In order that A and B$_1$ particles are bound, it is necessary that
\be
\dfrac{\sigma_{\rm A}+1/3}{2}\le\sqrt{\dfrac{a^2}{3}+c^2}\le\dfrac{\sigma_{\rm A}+1/3}{2}+\dfrac{1}{3}\,;
\label{eq:sheet1}
\ee
to prevent overlap between spheres and B$_2$ monomers, the condition is
\be
\sqrt{\dfrac{a^2}{3}+\left(c+\dfrac{2}{3}\right)^2}\ge\dfrac{\sigma_{\rm A}+1}{2}\,;
\label{eq:sheet2}
\ee
finally, we require that
\be
a\dfrac{\sqrt{3}}{3}\ge\sigma_{\rm B_2}=1
\label{eq:sheet3}
\ee
to rule out the possibility of lateral overlap between B$_2$ monomers. The value of $a$ is not exactly known but, from a glance at many snapshots, we expect that its typical value is only slightly larger than $\sigma_{\rm A}$. In Fig.~\ref{fig:theory} we analyze three cases (from left to right, $a/\sigma_{\rm A}=1$, 1.1, and 1.2): Eq.~(\ref{eq:sheet1}) is satisfied for $c$ values within the two red lines; Eq.~(\ref{eq:sheet2}) is satisfied for $c$ values above the blue line; Eq.~(\ref{eq:sheet3}) is satisfied for $\sigma_{\rm A}$ values falling on the right of the vertical black line; finally, $c\ge 1/6$ holds above the purple line. All in all, the only $\sigma_{\rm A}$ and $c$ values consistent with the existence of (flat, infinite) triangular membranes are those within the green regions in Fig.~\ref{fig:theory}. As $a/\sigma_{\rm A}$ increases, the range of $d=\sigma_{\rm A}/\sigma_{\rm B_2}$ where membranes exist gets reduced; in all three cases considered, $d=1$ is never allowed.

%
%

\section{CONCLUSIONS}

We have found a rich self-assembly behavior in a colloidal mixture of spheres and dimers with only a few basic assumptions on the interaction governing particle aggregation. Upon varying the sphere size and the relative concentration of the species, our system spontaneously gives rise to as many diverse aggregates as micelles, gel-like networks, bilayers, crystalline membranes, and vesicles, providing further proof of the stunning simplicity of the mechanism behind many complex structures also present in nature. Since colloidal particles with characteristics similar to those envisaged in the present model can actually be engineered, e.g. with the method explained in Ref.~\citenum{Wolters}, our results may readily find application in various fields, such as encapsulation technology or the development of novel mesoporous materials for heterogeneous catalysis. We are aware that the outcome of a simulation may in principle depend on the kinetics imposed by the algorithm chosen. This means that any discrepancy between the simulation dynamics and an experimental realization of the model on the colloidal scale could lead to differences in self-assembly products. For example, MC simulations with local moves only can have problems with equilibrating clustering systems. The use of cluster moves may alleviate this problem, see e.g. Ref.~\citenum{Whitelam}. In this regard, we observe that the use of a smart technique such as the aggregation-volume-bias algorithm does not change the self-assembly behavior of our system in any respect.~\cite{Prestipino1} In the near future, we plan to carry out a more refined exploration of the parameter space of the model, with the aim to characterize the boundaries between the various self-assembled structures and the mechanisms of crossover between them.

\section*{ACKNOWLEDGMENTS}
We dedicate this paper to the memory of our collaborator and dear friend Mimmo Gazzillo, theoretical physicist at Ca' Foscari in Venice, who unexpectedly passed away last year, aged 68.

This work has been done using the computer facilities made available by the PO-FESR 2007-2013 Project MedNETNA (Mediterranean Network for Emerging Nanomaterials).

\end{document}